**Metastability limit of pristine 2D noble metals with high-energy facet: dominance of the Bell–Evans–Polanyi principle**

Xiaoliang Zhong

xzhong@hust.edu.cn

School of Energy and Power Engineering, Huazhong University of Science and Technology, 1037 Luoyu Road, Wuhan, China

Using density functional theory, we predict that ultrathin (110) sheets of Rh, Pd, Ag, and Ir tend to transform into their (100) counterparts via lattice contraction. An approximately linear relationship between the transformation barrier and the energy difference between (110) and (100) sheets is revealed, demonstrating that the Bell–Evans–Polanyi (BEP) principle dominates. Furthermore, the critical thicknesses for these metals are also described by the BEP principle: below these thicknesses, the (110) sheets are no longer metastable and undergo spontaneous structural transformation.

Metastable materials can exhibit excellent properties that equilibrium-state materials cannot match. Recently, metal nanostructures with high-energy facets have attracted extensive research interest mainly because of their outstanding catalytic properties[1-4]. As the size of metal nanostructures decreases, the specific surface area increases, leading to a higher tendency of structural transformation into more stable structures and therefore face a fundamental trade-off between high performance and low structural stability.

When the thicknesses of metal nanosheets are small enough, the structures may transform into those with lower surface energies[5-7]. Concerning the stability of 2D noble metals, we herein address a fundamental problem that is not well understood, namely, how the energy barriers to more stable structures scale with sheet thickness. A lower energy barrier signifies a higher tendency of structural transformation; when the barrier vanishes, a spontaneous transformation is expected which in practice determines the synthesis limit of these metastable structures. Hasmy and Medina performed molecular dynamics simulations to estimate the energy barriers of the (001)-to-(111) transition in ultrathin Au films and found that an eight-atomic-layer film required a lower transformation temperature than thicker films[6]. However, a clear barrier - thickness relationship was still missing.

We investigate fcc noble metal (Rh, Pd, Ag, Ir, Pt and Au) nanosheets with high-energy (110) facets by applying DFT (density functional theory[8-9]) methods, for which major progress has recently been made in the synthesis[10-14]. Since the (110) surface energy of fcc metals is significantly higher than those of (111) and (100)[15], we study the potential spontaneous transformation of (110)-faceted nanosheets with decreasing thickness. The adopted models are freestanding ones with clean surfaces; therefore, we are studying the intrinsic metastability of the nanosheet. Once the metastability of the pristine nanosheets is well understood, we are also in a better position to elucidate the effects of factors such as substrates and adsorbates, both of which are known to significantly alter the stability of metal nanosheets[16-17].

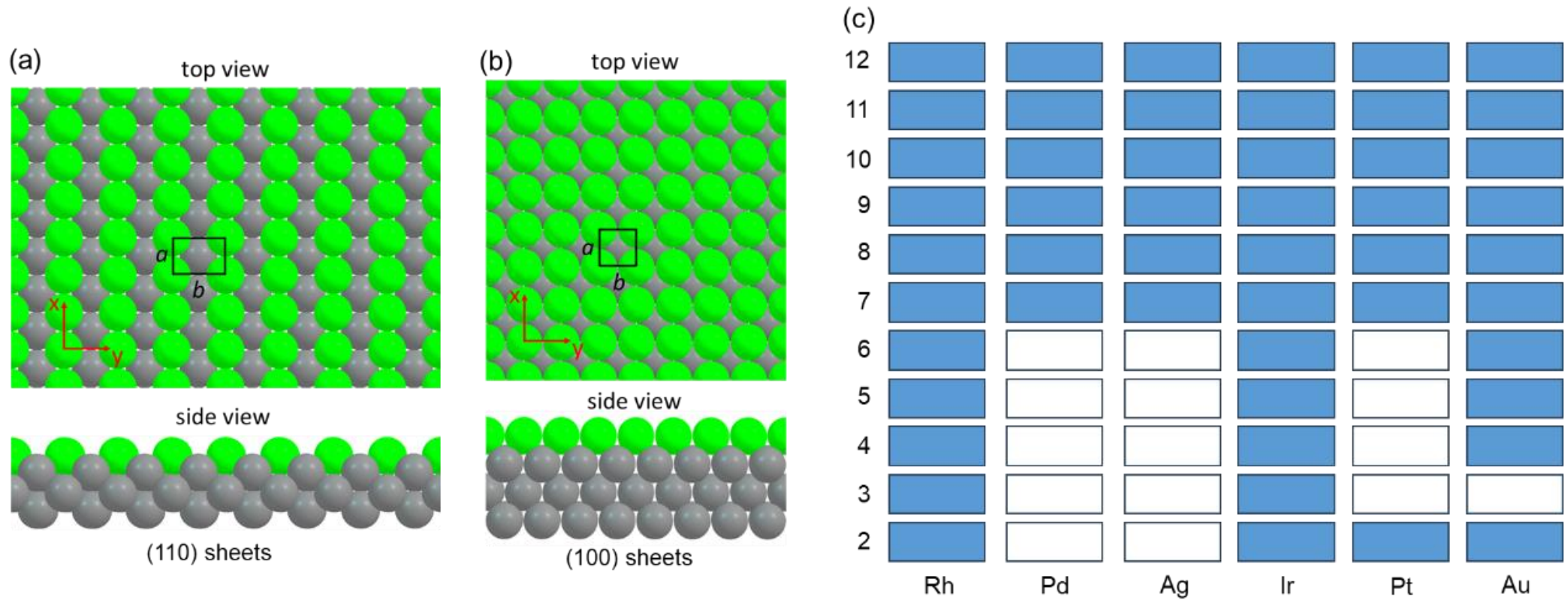

FIG. 1. Top view and side view of nanosheets exposing the (110) (a) and (100) (b) facets. The black rectangles represent the respective unit cells. The top atomic layer is in green for clarity. (c) Structural information of the initial (110) nanosheets. While the blue color denotes that the optimized structures retain the (110) facets, the white color means the structures turn into the (100) sheets after geometry optimization.

In the VASP calculations, the electron-ion interactions are described by the projector augmented plane wave (PAW) pseudopotentials. The plane-wave cutoff energy is set to 400 eV in all computations, which is larger than 1.3 times ENMAX in all POTCAR files. The convergence of energy is set as $10^{-6}$ eV. 16×16×1 and 16×8×1 Monkhorst–Pack k-point meshes are set for the unreconstructed and (1×2)-reconstructed (110) sheets, respectively. A vacuum layer of about 15Å along the c direction is added to avoid the interaction between periodic images.

We present in Fig. 1 the results of nanosheet geometry optimization with a thickness ranging from two to 12 monolayers (ML). These sheets initially take the bulk lattice parameters and expose the (110) surfaces (Fig. 1(a)). As shown in Fig. 1(c), the stiffer metals of Rh and Ir show a high robustness of the (110) facet; nanosheets of all thicknesses retain the (110) surfaces after structural relaxation. On the other hand, the more flexible metals of Pd, Ag, Pt and Au exhibit a lower robustness, that is, at some smaller thicknesses sheets transform into those with the (100) surfaces (Fig. 1(b)). This structural transformation is characterized by a decrease of ~30% in the lattice parameter *b* accompanied by an increase in the nanosheet thickness.

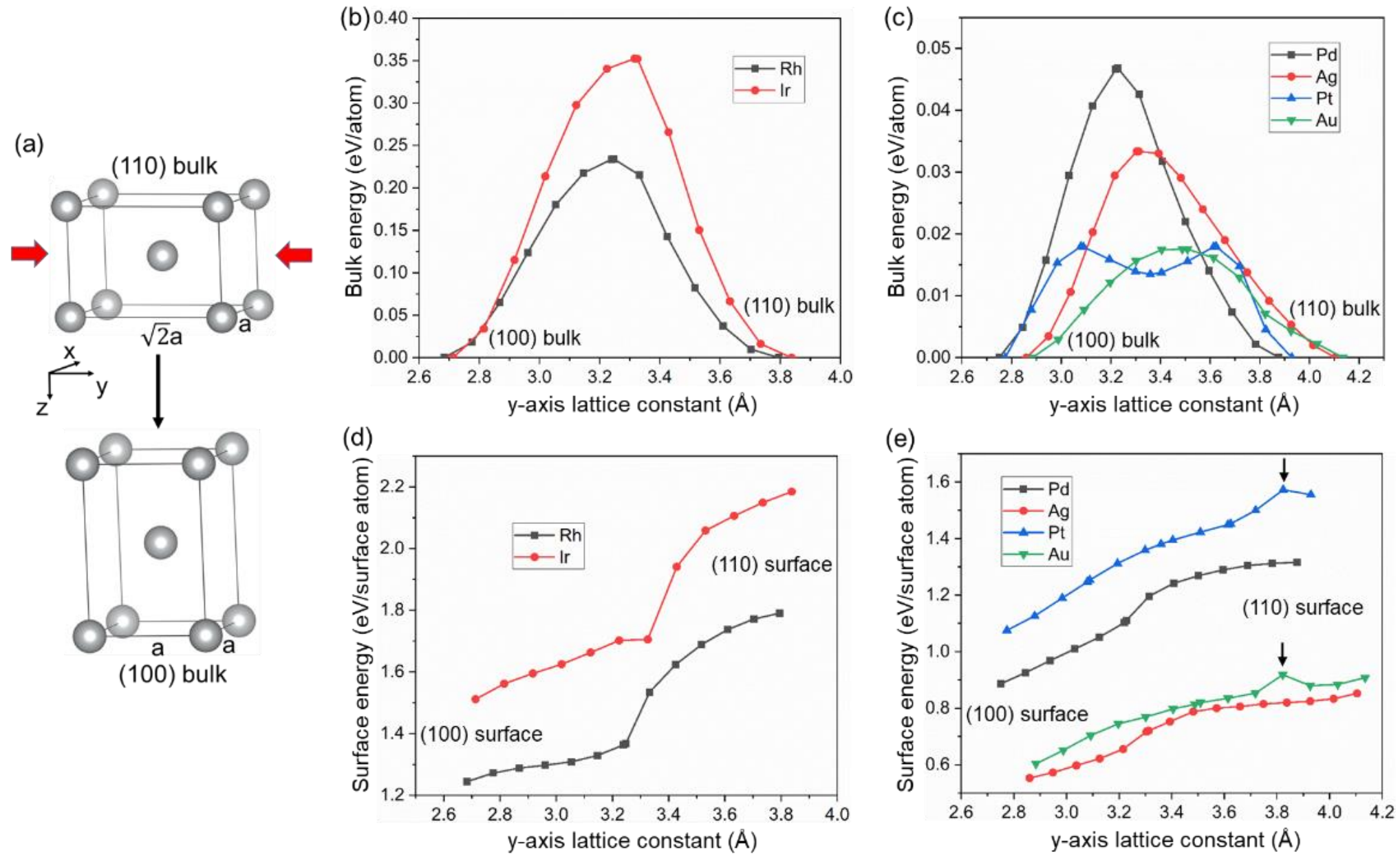

FIG. 2. (a) Schematic view of the bulk phase transformation. Bulk energy variation with lattice constant is shown in (b) and (c), while surface energy variation is shown in (d) and (e).

The spontaneous phase transformation is driven by the reduction of the sheet total energy without encountering an energy barrier ($E_b$). Structurally, one metal nanosheet can be decomposed into a bulk region and two surface regions. After the structural transition, the bulk region retains the FCC structure (Fig. 2(a)). Therefore, energy reduction originates from the lower surface energy of the (100) facet compared with the (110) one. Throughout this manuscript, surface energies are given in unit of eV per surface atom, since the number of surface atoms rather than surface areas is unchanged after the transformation. As Fig. 2(a) shows, during the structural transition lattice parameter *b* decreases from $\sqrt{2}$a to a while lattice parameter *a* remains unchanged (=a) before and after the transition. Meanwhile, lattice parameter *c* increases from a to $\sqrt{2}$a, resulting in an increase of the sheet thickness (Fig. 1). The energy evolution of the bulk region as shown in Fig. 2 is calculated by fixing a set of lattice parameter *b* while *a* and *c* are optimized automatically. For each metal species energy of the '(100) bulk' (the left end) is the same as the '(100) bulk' (the right end) since they correspond to the same FCC crystal just with different orientations (see Fig. 2(a)).

Relating with the high stiffness, both Rh and Ir exhibit a high energy barrier in this structural transition. The resolution of $b$-variation is reduced to 0.01 Å around energy maxima to obtain the heights of $E_b$. The four more flexible metals on the other hand show much lower energy barriers (Fig. 2(c)). Interestingly, in the case of Pt there is an energy minimum at $b = 3.359$ Å, suggesting that a metastable state may exist therein. This in consistent with a previous DFT study, which shows among various metals only Pt has a stable body-centered tetragonal (BCT) phase[18].

The evolution of surface energies during the transition is also shown in Fig. 2. Similar to bulk energy variation, the stiffer metals (Rh and Ir) show higher surface energies for both the (100) facet (the left ends) and the (110) facet (the right ends). For Rh, Ir, Pd and Ag, surface energies decrease monotonically with the decrease of lattice parameter $b$ until the surface turns into the (100) facet. Therefore, for these species there two competing factors during the transformation from the (110) nanosheets to the (100) ones. The first one is the energy barrier of the bulk region which resists the transformation. The other one is the monotonic decrease of surface energy with decreased $b$ which favors the transformation. For thicker sheets the bulk region dominates and the energy barrier of this region keeps the (110) ones from transforming into the (100) ones spontaneously. For very thin nanosheets the surface effects may dominate which makes the transformation spontaneous. In the case of Pt and Au, there exist a local maximum in surface energy as denoted by the black arrows in Fig. 2(e). On the right side of the local maximum both bulk energy and surface energy increase with the decrease of $b$.

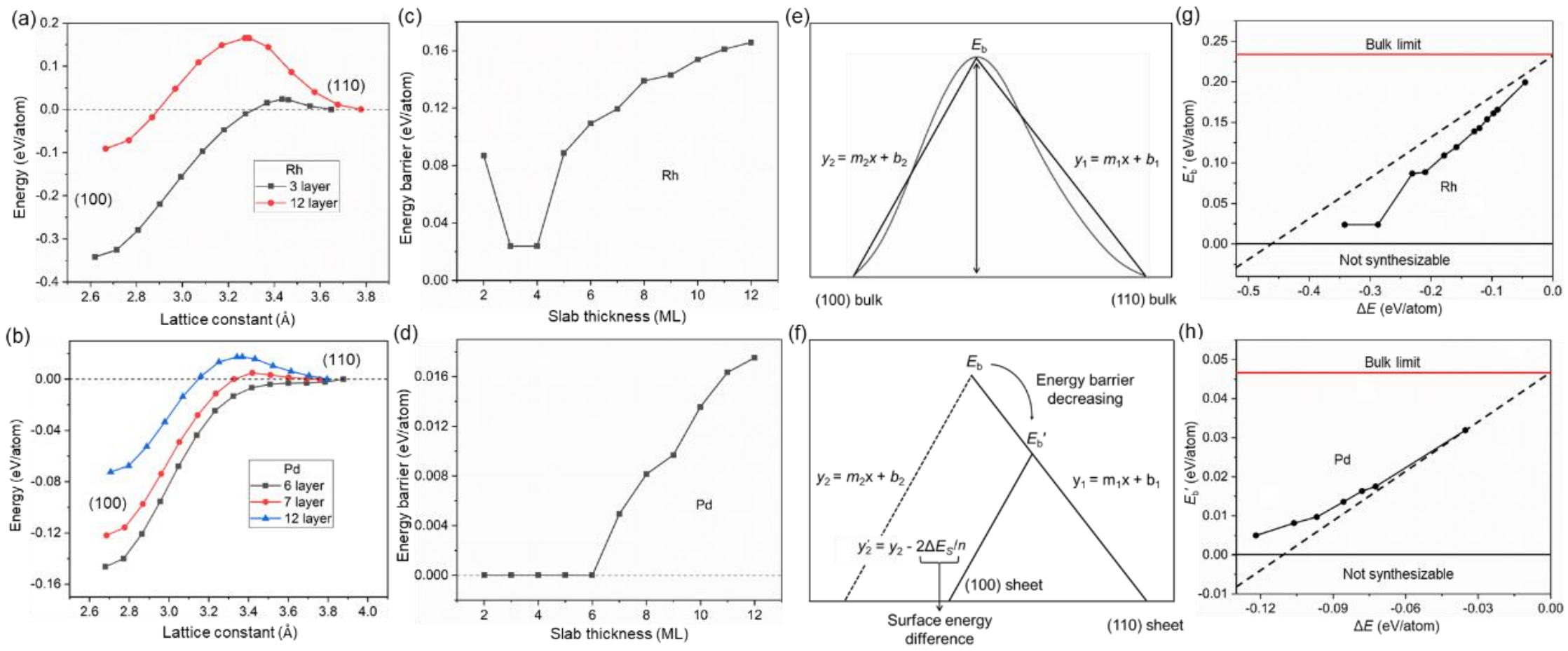

FIG. 3. Energy evolution of nanosheets with different thicknesses for Rh (a) and Pd (b). The corresponding energy barriers with different thicknesses are shown in (c) and (d). (c) The actual bulk energy variation is approximated by two straight lines. (d) The lowering of the left line in a sheet problem. (g)-(h) The black solid lines representing the actual energy barriers for Rh (g) and Pd (h) as a function of sheet thickness while the black dashed ones representing the derived barriers based on the BEP relationship. The red lines show the corresponding energy barriers of the bulks.

To test the validity of this idea, we plot in Fig. 3 the total energy variation of the Rh and Pd nanosheets as a function of lattice parameter $b$. At a greater thickness of 12 ML, Rh and Pd exhibit a single energy barrier of 0.166 (Fig. 3(a) red line) and 0.018 eV/atom (Fig. 3(b) blue line), respectively. When the thickness is reduced to 7 ML, $E_b$ of Pd decreases to 0.005 eV/atom, showing increased surface effects. At an even smaller thickness of 6 ML, energy barrier vanishes (Fig. 3(b) black line). The thickness-dependent energy barrier of Rh and Pd is plotted in Fig. 3 (c) and (d). With the decrease of thickness $E_b$ in general decreases, and in the case of Pd, $E_b$ vanishes for the thickness from 6 to 2 ML. This in fact means that for Pd, (110) nanosheets with smaller thicknesses are *not* synthesizable; in turn, they would spontaneously transform to the (100) ones (Fig. 1). On the other hand, due to the larger energy barrier of the bulk region (Fig. 2), Rh exhibits a finite $E_b$ down to a thickness of 2 ML. Therefore, Rh nanosheets with all thicknesses retain the (110) facet after the geometry relaxation (Fig. 1).

We propose that the variation of energy barrier with sheet thickness is analogy to the classic Bell–Evans–Polanyi (BEP) principle, which states that reaction activation energy has a linear relationship with enthalpy of reaction. Regarding metal nanosheets, structural transformation energy barrier is analogous to activation energy while energy difference between a (110) sheet and a (100) one is analogous to enthalpy of reaction. When the BEP principle is invoked, the two energy functions between the reactant, the transition state, and the product are approximated by two straight lines. For metal bulks, we also approximate energy evolution (see Fig. 2(b) and (c)) by two straight lines with the intersection denoting the energy barrier (Fig. 3 (e)). The equations of the two straight lines are

$$y_1 = m_1 x + b_1 \tag{1}$$

and

$$y_2 = m_2 x + b_2 \tag{2}$$

Then the energy barrier (the height of the intersection point) is

$$E_{\mathrm{b}} = \frac{m_1 b_2 - m_2 b_1}{m_1 - m_2} \tag{3}$$

With respect to the nanosheets, surface energies need to be considered. Compared with a (110) sheet, the (100) one with the same thickness (in ML) has a lower energy originating from the lower surface energy. If we formally still use equation (1) to approximate the energy variation from the (110) sheet to the intersection point, the straight line from the (100) sheet to the intersection point becomes (Fig. 3 (f))

$$y_2' = y_2 - \frac{2\Delta E_S}{n} \tag{4}$$

where $\Delta E_S$ is the surface energy difference (in units of eV/surface atom) between the (110) and (100) facets while $n$ is the sheet thickness in ML. The factor of two accounts for the two surfaces of a metal sheet. The presence of the denominator $n$ is because the total energies are always given in units of eV/atom for comparing sheets with different thicknesses. The intersection point of the two straight lines as given by equation (1) and equation (4) (see Fig. 3 (f)) denotes the location of energy barrier of the nanosheets,

$$E_b' = E_{\mathrm{b}} - \frac{2m_1}{(m_1 - m_2)} \frac{\Delta E_S}{n} \tag{5}$$

In Fig. 3 (g) and (h) we plot the calculated energy barriers (Fig .2 d-f) versus the derived ones (equation (5)) as a function of $\Delta E_S$. A larger thickness of 24 ML is also included to test our hypothesis. It appears that the two sets of barriers in general agree reasonably well with each other especially for thick sheets. For very thick sheets energy barriers should converge to the corresponding bulk values since surface effects can be neglected therein. That is, as equation (5) shows, when $n \to \infty$ the second term on the right-hand side vanishes therefore $E_b' = E_{\mathrm{b}}$. The critical thickness $n^*$ can be obtained by solving equation (5) for $E_b' = 0$

$$n^* = \frac{2m_1}{m_1 - m_2} \frac{\Delta E_S}{E_b} \tag{6}$$

With the calculated parameters as shown in Table 1, the derived critical thicknesses are 2.4, 7.8 and 6.5 ML for Rh, Pd and Ag, respectively. Previously, by fully optimizing the sheet structures, the actual critical thicknesses are shown to be 7 ML for Pd and 5 ML for Ag, respectively, while Rh (110) sheets are always stable against a spontaneous transformation down to 2 ML (Fig. 1). Therefore, it shows that the classic BEP principle plays a dominate role in determining both the variation in energy barrier and the critical thickness.

**Table 1 | Relevant parameters used in the equations**

| | m1 | m2 | b1 | b2 | $E_{\mathrm{b}}$ | $\Delta E_S$ |
|---|---|---|---|---|---|---|
| Rh | -0.427 | 0.414 | 1.621 | -1.111 | 0.234 | 0.546 |
| Pd | -0.072 | 0.098 | 0.280 | -0.269 | 0.047 | 0.430 |
| Ag | -0.042 | 0.074 | 0.173 | -0.211 | 0.033 | 0.299 |

In this work we have demonstrated that as thickness decreases, the tendency of 4d noble metal (110) nanosheets to transform increases and finally the metastable (110) sheets can become unattainable. The underlying physics can be universal to nanosheets with higher-energy facets, that is, when the specific surface area increases, the bulk region is no longer able to provide sufficient energy barrier to structural transformation. We have also shown that when the (110) sheets are compressed along the y direction, energy barrier decreases as thickness decreases in a manner dominated by the BEP principle. Accordingly, the critical thicknesses of Pd and Ag with vanishing energy barriers are also dominated by the BEP principle. Therefore, we have extended the application of the classic BEP principle to the study of transformation energy barriers and synthesizability of low dimensional materials.